\documentclass[aps,pra,twocolumn,10pt,longbibliography,superscriptaddress]{revtex4-2}

\usepackage[usenames]{xcolor}
\usepackage[hidelinks]{hyperref}

\usepackage[english]{babel}
\usepackage[utf8]{inputenc} 
\usepackage[T1]{fontenc} 
\usepackage{times}

\usepackage{amsthm} 
\usepackage{amssymb} 
\usepackage{amsmath} 
\usepackage{mathbbol} 
\usepackage[normalem]{ulem}
\usepackage{graphicx}
\usepackage{paralist} 

\newtheorem{Algorithm}{Algorithm}
\newtheorem{Problem}{Problem}

\usepackage{mathtools} 

\makeatletter
\newcommand{\manuallabel}[2]{\def\@currentlabel{#2}\label{#1}}
\makeatother

\makeatletter
\newcommand{\pushright}[1]{\ifmeasuring@#1\else\omit\hfill$\displaystyle#1$\fi\ignorespaces}
\makeatother

\renewcommand{\i}{\mathrm{i}}
\newcommand{\1}{\mathbb{1}}

\DeclareMathOperator{\Tr}{Tr}

\newcommand{\KetBra}[1]{{\Ket{#1}\!\Bra{#1} }}

\newcommand{\Bra}[1]{{ \langle \! \langle{#1}\vert }}
\newcommand{\Ket}[1]{{ \vert {#1}  \rangle \!  \rangle}}

\newcommand{\ket}[1]{\left|{#1}\right\rangle}

\newcommand{\bra}[1]{\left\langle{#1}\right|}

\newcommand{\ketbra}[2]{\ket{#1} \!\! \bra{#2}}

\newcommand{\id}{\mathbb{1}}

\usepackage{multirow}

\begin{document}

\title{Computational advantage from quantum superposition of multiple temporal orders of photonic gates}

\author{M\'arcio M.\ Taddei}
\email{marciotaddei [at] gmail.com}
\affiliation{Instituto de F\'isica, Federal University of Rio de Janeiro, 21941-972, P.\ O.\ Box 68528, Rio de Janeiro, Brazil}
\affiliation{ICFO - Institut de Ciencies Fot\`oniques, The Barcelona Institute of Science and Technology, 08860, Castelldefels, Barcelona, Spain}

\author{Jaime Cari\~ne}
\affiliation{Departamento de F\'{\i}sica, Universidad de Concepci\'on, 160-C Concepci\'on, Chile}
\affiliation{ANID – Millennium Science Initiative Program – Millennium Institute for Research in Optics, Universidad de Concepci\'on, 160-C Concepci\'on, Chile}
\affiliation{Departamento de Ingenier\'ia El\'ectrica, Universidad Cat\'olica de la Sant\'isima Concepci\'on, Alonso de Ribera 2850, Concepci\'on, Chile}

\author{Daniel Mart\'inez}
\affiliation{Departamento de F\'{\i}sica, Universidad de Concepci\'on, 160-C Concepci\'on, Chile}
\affiliation{ANID – Millennium Science Initiative Program – Millennium Institute for Research in Optics, Universidad de Concepci\'on, 160-C Concepci\'on, Chile}

\author{Tania Garc\'ia}
\affiliation{Departamento de F\'{\i}sica, Universidad de Concepci\'on, 160-C Concepci\'on, Chile}
\affiliation{ANID – Millennium Science Initiative Program – Millennium Institute for Research in Optics, Universidad de Concepci\'on, 160-C Concepci\'on, Chile}

\author{Nayda Guerrero}
\affiliation{Departamento de F\'{\i}sica, Universidad de Concepci\'on, 160-C Concepci\'on, Chile}
\affiliation{ANID – Millennium Science Initiative Program – Millennium Institute for Research in Optics, Universidad de Concepci\'on, 160-C Concepci\'on, Chile}
 
\author{Alastair A.\ Abbott}
\affiliation{Department of Applied Physics, University of Geneva, 1211 Geneva, Switzerland}
\affiliation{Univ. Grenoble Alpes, Inria, 38000 Grenoble, France}

\author{Mateus Ara\'ujo}
\affiliation{Institute for Quantum Optics and Quantum Information (IQOQI), Austrian Academy of Sciences, Boltzmanngasse 3, 1090 Vienna, Austria}

\author{Cyril Branciard}
\affiliation{Univ.\ Grenoble Alpes, CNRS, Grenoble INP, Institut N\'eel, 38000 Grenoble, France}

\author{Esteban S.\ G\'omez}
\affiliation{Departamento de F\'{\i}sica, Universidad de Concepci\'on, 160-C Concepci\'on, Chile}

\author{Stephen P.\ Walborn}
\affiliation{Departamento de F\'{\i}sica, Universidad de Concepci\'on, 160-C Concepci\'on, Chile}
\affiliation{ANID – Millennium Science Initiative Program – Millennium Institute for Research in Optics, Universidad de Concepci\'on, 160-C Concepci\'on, Chile}

\author{Leandro Aolita}
\affiliation{Instituto de F\'isica, Federal University of Rio de Janeiro, 21941-972, P.\ O.\ Box 68528, Rio de Janeiro, Brazil}
\affiliation{Quantum Research Centre, Technology Innovation Institute, Abu Dhabi, UAE}

\author{Gustavo Lima}
\affiliation{Departamento de F\'{\i}sica, Universidad de Concepci\'on, 160-C Concepci\'on, Chile}
\affiliation{ANID – Millennium Science Initiative Program – Millennium Institute for Research in Optics, Universidad de Concepci\'on, 160-C Concepci\'on, Chile}

\begin{abstract}
Models for quantum computation with circuit connections subject to the quantum superposition principle have been recently proposed. There, a control quantum system can coherently determine the order in which a target quantum system undergoes $N$ gate operations. This process, known as the quantum $N$-switch, is a resource for several information-processing tasks. In particular, it provides a computational advantage --- over fixed-gate-order quantum circuits --- for phase-estimation problems involving $N$ unknown unitary gates. However, the corresponding algorithm requires an experimentally unfeasible target-system dimension (super-)exponential in $N$. Here, we introduce a promise problem for which the quantum $N$-switch gives an equivalent computational speed-up with target-system dimension as small as 2 regardless of $N$. We use state-of-the-art multi-core optical-fiber technology to experimentally demonstrate the quantum $N$-switch with $N$=4 gates acting on a photonic-polarization qubit. This is the first observation of a quantum superposition of more than $N$=2 temporal orders, demonstrating its usefulness for efficient phase-estimation.
\end{abstract}
\maketitle

\section{Introduction}

Quantum mechanics allows for processes where two or more events take place in a quantum superposition of different temporal orders. This exotic phenomenon results in causal nonseparability \cite{Oreshkov2012,Araujo2015,Oreshkov2016}, and it is likely to be especially relevant in quantum treatments of gravity \cite{Hardy2005,Hardy2007,Hardy2009}. In fact, quantum control of temporal orders could be realized with quantum circuits exploiting hypothetical closed time-like curves \cite{Chiribella2013,Araujo2017a}, and it would also arise naturally due to the spacetime warping that macroscopic spatial superpositions of massive bodies would cause \cite{Zych2019}.

From a more practical perspective, advanced quantum computational models without definite gate orders have sparked a great deal of fundamental interest, as they do not fit into the usual paradigm of circuits with fixed gate connections \cite{Chiribella2008,Hardy2009,Chiribella2009,Chiribella2012,Colnaghi2012, Chiribella2013}. The best known example is the celebrated quantum $N$-switch gate, $S_N$, which coherently applies a different permutation of $N$ given gates on a target quantum system conditioned on the state of a control quantum system \cite{Chiribella2013,Colnaghi2012,Araujo2014}. $S_N$ has been identified as a resource for a number of exciting information-theoretic tasks. For instance, for $N$$=$$2$, it allows one to deterministically distinguish pairs of commuting versus anti-commuting unitaries \cite{Chiribella2012}; and, remarkably, this translates into an exponential advantage in a communication complexity problem \cite{Guerin2016, Wei2019}.

In general, circuits that synthesize $S_N$ with a fixed gate order are known, but at the expense of quadratically more queries to (i.e., uses of) the gates \cite{Chiribella2012,Colnaghi2012, Araujo2014, Facchini2014}. As a consequence thereof, $S_N$ allows one to solve a promise problem \cite{Chiribella2012,Araujo2014} on the permutations of $N$ unknown unitary gates with quadratically fewer queries in $N$ than all known circuits with fixed gate order. More precisely, the permutation sequences of the gates are promised to differ only by a phase factor, and $S_N$ efficiently estimates these phase differences.  However, the algorithm for this problem \cite{Chiribella2012,Araujo2014} requires the target-system dimension to grow (super-)exponentially with $N$, making it experimentally demanding. As a matter of fact, all experimental realizations of the quantum $N$-switch reported so far are restricted to the simplest case of $N$$=$$2$ gate orders \cite{Procopio2014,Rubino2016,Goswami2018,Wei2019,Guo2020,Goswami2018a}.

\begin{figure*}[t!]
\centering
\includegraphics[width=2\columnwidth]{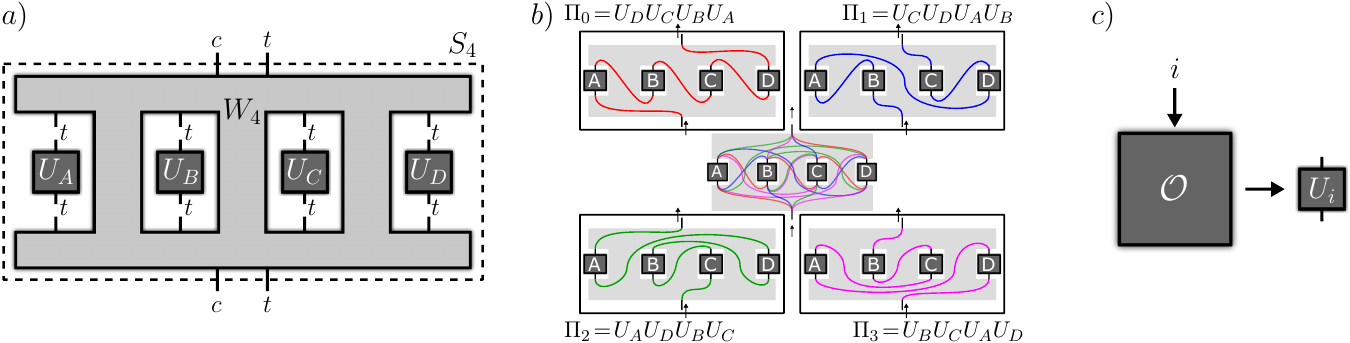}
\caption{ 
$a$) Abstract representation of the quantum $N$-switch for the case of $N=4$. The process, $W_4$ (light-grey region), can be thought of as an experimental setup (e.g., a quantum circuit or interferometer) through which the composite control-target system goes and with open slots for target-subsystem gates $U_i$ (dark-grey boxes), for $i=A$, $B$, $C$, or $D$, to be inserted. Inside $W_4$, the connections between these gates are coherently controlled by the control subsystem, an effect known as quantum control of gate orders (QCGO). This property is a physical resource for certain quantum computations (phase-estimation problems), and $W_4$ is the resourceful object that bears it. The concatenation of $W_4$ with the inserted gates yields the quantum 4-switch gate $S_4$, a joint unitary operation on the composite system. 
$b$) Concrete schematics of the specific variant of the quantum 4-switch process experimentally implemented in this work. The target subsystem undergoes the four-gate sequence in a quantum superposition (center) of $P=4$ different orderings (permutations of the string $ABCD$): $ABCD, BADC, CBDA, DACB$. Each permutation is shown individually in a different color and panel.
$c$) In the above-mentioned computations, the target-subsystem gates are unknown. For the purpose of complexity analysis, they can be thought of as produced upon request by a quantum oracle $\mathcal{O}$. This takes as input $i=A$, $B$, $C$, or $D$ and outputs a black-box device implementing the unknown gate $U_i$. Each such call to the oracle counts as an oracle query. The $N$-switch process allows one to solve computational problems on the phase relationships between permutations of the black-box gates with considerably fewer oracle queries -- i.e.\ lower query complexity -- than any process with fixed (or classically controlled) gate connections.
}
\label{fig:4switchprocess}
\end{figure*}

In this work, we introduce a novel algorithm that exploits the quantum $N$-switch and experimentally demonstrate it for $N$$=$$4$ unitary gates. Specifically, we find a variant of the above phase-estimation problem, which we name the Hadamard promise problem, for which the quantum $N$-switch is also a resource but with considerably milder constraints on the target-system dimension. 
On one hand, this problem plays a role in computation with indefinite gate orders analogous to Deutsch-Jozsa's \cite{Deutsch1992} or Simon's \cite{Simon1997} problems in the beginnings of quantum computation: a proof-of-principle of improvements over a previous paradigm. On the other hand, there are reasons to expect that practical applications of the Hadamard promise problem will be developed, both because closely related phase-estimation problems already have many applications, and because it involves the quantum Fourier transform, which is an important sub-routine for a variety of quantum algorithms with practical applications \cite{Nielsen2010}.
The problem's promise is that the products of the $N$ unknown gates applied in $P$ different orders differ only in $+$ or $-$ signs that are encoded into one of the columns of a given $P{\times}P$-dimensional Hadamard matrix; and the problem consists of finding which column it is.

The algorithm to solve this problem exploits the quantum $N$-switch -- consuming $N$ queries to the gates -- to deterministically find the column. This represents a speed-up quadratic in $N$ in query complexity (i.e.\ number of queries) with respect to all known algorithms exploiting circuits with fixed gate orders (see \cite{Araujo2014,Friis2014,Oreshkov2019} for a discussion of how to count queries in a quantum switch). Hence, the algorithm is not only an interesting computational primitive on its own but also a practical tool to benchmark experimental realizations of $S_N$, because the quantum $N$-switch is the only known process for which the algorithm succeeds with unit probability for all gates satisfying the promise while only consuming $N$ gate queries. To demonstrate the practicability of the algorithm we implement it with a quantum $N$-switch of $N=4$ gates using modern multi-core optical-fiber technology \cite{Richardson2013,Canas2017,Ding2017,Xavier2020}.  The 4 gates are implemented on the target polarization qubits using programmable liquid-crystal devices, and the spatial degree of freedom of a single photon is used as the control system. We obtain an average success probability for the algorithm, over different sets of gates, of $p_\text{succ}\approx 0.95$. Our results represent the first demonstration of the quantum $N$-switch gate for $N$ larger than 2, as well as of its efficiency for phase estimation problems involving multiple unknown gates.

\section{Preliminaries}
\subsection*{Quantum control of gate orders}
\label{sec:prelim1}   
In quantum computation, a quantum switch can be described by a special type of controlled operation that applies a particular unitary gate $\Pi_x$ to a target system ($t$) for each different state of a control system ($c$).  We define the quantum $N$-switch gate as
\begin{equation}
S_N \ket{x}_c\ket{\Psi}_t=  \ket{x}_c\, \Pi_x \ket{\Psi}_t \ ,
\label{eq:def_SN}
\end{equation}
where $\ket{x}_c$ is the $x$-th member of the computational basis of the control system, and $\ket{\Psi}_t$ is an arbitrary state of the target system. The heart of the quantum $N$-switch is the operator $\Pi_x\coloneqq U_{\sigma_x(N-1)}\hdots U_{\sigma_x(1)}U_{\sigma_x(0)}$, which is a product of the $N$ unitary gates in a fixed set  $\mathsf U \coloneqq\{U_A,U_B,...\}$, in their $x$-th ordering. More precisely, $\sigma_x$ is a vector  with $N$ elements specifying the $x$-th permutation of the $N$ gates in $\mathsf U$, i.e. it specifies the ordering sequence of the unitaries, so that  $\sigma_x(j)$ is the $j$-th element in the $x$-th permutation. To control the implementation of $P$ different permutations of gates requires a control system of at least dimension $P$.  The dimension of the target system can be arbitrary and we denote it as $d$.    With $S_N$ defined as in \eqref{eq:def_SN}, it is clear that $c$ coherently controls the order of the $N$ unitary gates applied to system $t$, which explains the name ``quantum control of gate orders'' (QCGO).
We note that the usual definition  \cite{Colnaghi2012,Araujo2014} of the quantum $N$-switch deals only with the specific case of all $N!$ permutations of the gates in $\mathsf U$. However, here (as in Refs.\cite{Procopio2019,Procopio2020}) we will be interested in the more general case $P\leq N!$.

Clearly, the general definition of QCGO is independent of the specific choice of gates in $\mathsf U$. A convenient mathematical tool to capture that is the quantum $N$-switch process $W_N$, which produces the quantum $N$-switch gate $S_N$ when given the set of gates $ \mathsf U $ as input. For the technical definition of processes, we refer the reader to Refs. \cite{Oreshkov2012,Araujo2015,Oreshkov2016,Araujo2017}. 
Intuitively, one can think of a process as the quantum evolution generated by an experimental arrangement with open slots for gates on the target system to be inserted \cite{Chiribella2008,Chiribella2009}, as represented in Fig.~\ref{fig:4switchprocess}~(a). Inside the process, the connections between the inserted gates may be subject to the quantum superposition principle. 
For instance, Fig.~\ref{fig:4switchprocess} (b) pictorially represents our experimental implementation of the quantum $4$-switch $S_4$, with a coherent quantum superposition of $P=4$ different gate connections (each one in a different color), for the particular choice of permutation set $\{ABCD, BADC, CBDA, DACB\}$.  
Such superpositions give rise to QCGO, which corresponds to a specific type of quantum control of causal orders\cite{Taddei2019} (and both phenomena are in turn contained within the general notion of causal nonseparability \cite{Oreshkov2012,Araujo2015,Oreshkov2016}).
In particular, QCGO takes place when those gate connections are coherently controlled by a control system, as in Eq.~\eqref{eq:def_SN}. Aside from being a fundamentally interesting phenomenon, QCGO turns out to be a physical resource for interesting phase-estimation problems, as we discuss next.

\subsection*{The Ara\'ujo-Costa-Brukner algorithm}
\label{sec:prelim2}
The quantum $N$-switch process provides an advantage for solving a particular phase-estimation problem \cite{Chiribella2012,Araujo2014} to which we here refer as the Fourier promise problem. In this type of problems, one has access to a quantum oracle $\mathcal{O}$ for $\mathsf U$, i.e.\ a black-box device that delivers a gate $U_i\in\mathsf U$ every time it is queried. See Fig.~\ref{fig:4switchprocess} (c). No information about the gates is available except for the promise that, for the constant phase factor $\omega\coloneqq e^{i \frac{2\pi}{P}}$ and for all $x\in[P]$, they satisfy the property:
\begin{equation}
\Pi_x = \omega^{x y}\, \Pi_0,
\label{eq:P_y}
\end{equation}
for some fixed, unknown $y\in[P]$, where the short-hand notation $[P]\coloneqq\{0, 1\hdots, P-1\}$ has been introduced. The task is to determine which one of the properties holds, i.e.\ to find $y$.

The Ara\'ujo-Costa-Brukner algorithm to solve this problem is based on the standard Hadamard test \cite{Hadamard}, and shares similarities with the Kitaev phase estimation algorithm \cite{Kitaev1995}. The control system is initialized in the computational-basis reference state $\ket{0}_c$, while the target system starts in an arbitrary state $\ket{\Psi}_t$. A $P$-dimensional quantum Fourier transform $F_{P}$ on $c$ maps it to a uniform superposition of all computational-basis states. Then, the quantum $N$-switch gate is applied. Because of property \eqref{eq:P_y}, this introduces the phase factor $\omega^{x y}$ to each computational-basis state $\ket{x}_c$ in the superposition, while the state $\Pi_0\ket{\Psi}_t$ of the target system factorizes. The value of $y$ is thus encoded into the phases of the superposition state of the control system. To map it back to the computational basis, one uncomputes the Fourier transform (applying its inverse $F_{P}^{-1} = F_P^\dagger$). In symbols \cite{Araujo2014}:
\begin{equation}
F_{P}^{-1}\,S_N\,F_{P} \ket{0}_c\, \ket{\Psi}_t= \ket{y}_c\, \Pi_0\ket{\Psi}_t\ .
\label{eq:invFourier}
\end{equation}
Then, $y$ is finally read out by a single-shot computational-basis measurement on $c$.

To apply $S_N$, one must consume $N$ queries to $\mathcal{O}$. Therefore, the query complexity -- i.e.\ total number of oracle queries -- of the algorithm is $Q=N$, for all $P\leq N!$. Remarkably, causally ordered processes (i.e., those produced by circuits with fixed, or classically controlled, gate connections) require considerably more queries to solve the same problem. For instance, for $P=N!$, the best causally ordered process displays query complexity $Q=\Omega(N^2)$ \cite{Colnaghi2012, Araujo2014, Facchini2014}, i.e.\ quadratically higher in $N$.
A downside of the algorithm, however, is that the target-system dimension $d$ must grow with the number $P$ of gate orders. This can be seen\cite{Araujo2014} by taking the determinant of both sides of Eq.~\eqref{eq:P_y}. For $y=1$, and since $\det \Pi_x = \det \Pi_0$, this imposes $\det \Pi_0 = \omega^{xd} \det \Pi_0$ (and, hence, $1=e^{i \frac{2\pi}{P} x d}$), for all $x\in[P]$, which is possible only if $d\ge P$.
This constraint is especially significant for experimental realizations, where coherently manipulating high-dimensional target systems together with high-dimensional control systems is challenging \cite{Wei2019}. For example, this limitation implies that if the polarization of a single photon ($d=2$) is used as the target system, the algorithm is useful only for $P=2$; despite the fact that the  spatial degree of freedom of the photon is amenable to encode much higher-dimensional control systems \cite{Aguilar2018}. To overcome this, we next introduce another variant of phase-estimation problem that is considerably less sensitive to the determinant constraint.

\section{A new computational primitive:\\the Hadamard promise problem}
\label{sec:new_algorithm}
We consider a different promise on the gates that the oracle $\mathcal{O}$ outputs.
Given a known $P{\times}P$-dimensional square matrix $M_P$ of entries $m_{x,y}=\pm1$, we require that the black-box unitaries in $ \mathsf U$ satisfy, for all $x\in[P]$, the property:
\begin{equation}
\Pi_x = m_{x,y} \ \Pi_0 \ ,
\label{eq:property_Hadamard}
\end{equation}
for some fixed, \emph{a priori} unknown matrix column $y\in[P]$. The task is, again, to find out $y$. In contrast to the complex-phase relation of Eq.~\eqref{eq:P_y}, the constraint that this real-phase relation imposes on $d$ is much softer. As one can see taking the determinant of both sides of Eq.~\eqref{eq:property_Hadamard}, the only requirement that arises now is that $(m_{x,y})^{d}=1$ for all $x,y\in[P]$, which is satisfied by any even $d$. With this, the promise problem finds application even when the target system is a simple qubit, regardless of the number of permutations $P$. Instead of a single complex phase factor, the value of $y$ is now encoded in a string of $P$ real phase factors (i.e., a column of $M_P$). The question, then, is how to decode that information. Luckily, the value of $y$ can be mapped back onto the computational basis of $c$ with a simple procedure, similar to that in Eq.~\eqref{eq:invFourier}, provided that $M_P$ is a \emph{Hadamard matrix} \cite{Hadamard}.

A Hadamard matrix (of order $P$) is a $P{\times}P$-dimensional square matrix $M_P$ with entries $m_{x,y}=\pm1$ and whose columns (or equivalently, whose rows) are all mutually orthogonal. The transpose $M_P^{\mathsf T}$ of $M_P$ is proportional to its inverse: $\frac{1}{P}M_P \cdot M_P^{\mathsf T}=\mathbb1$, with $\mathbb1$ the identity matrix. Such matrices can only exist for $P$ equal to 1, 2 or integer multiples of 4, and are conjectured to exist for all such dimensions. In fact, they can be generated recursively for any $P=2^k$, with $k\in\mathbb{N}$. 
Here we are actually interested in the subset of Hadamard matrices with all $+1$'s in the first row ($x=0$) and column ($y=0$). The former condition is required by Eq.~\eqref{eq:property_Hadamard}, whereas the latter condition is necessary in our algorithm below for correct encoding (see App.~\ref{subsec:proof_of_algorithm} for details).
With this, we can formally rephrase this promise problem as follows.

\begin{Problem}[Hadamard promise problem]
\label{prblm:Hadamard_permutation_problem}
Given a Hadamard matrix $M_P$ with all $+1$ entries along its first row and column and a unitary-gate oracle $\mathcal{O}$ fulfilling the promise -- i.e.\ Eq.~\eqref{eq:property_Hadamard} for some column $y\in[P]$ of $M_P$ --, compute $y$.
\end{Problem}

The algorithm to solve it with the quantum $N$-switch gate is similar to the Ara\'ujo-Costa-Brukner algorithm but with the quantum Hadamard gate $H_{P}$ associated to $M_P$ playing the role of $F_P$. The matrix representation of $H_{P}$ in the computational basis is $H_{P}\coloneqq \frac{M_P}{\sqrt{P}}$. Then, the following algorithm solves Problem \ref{prblm:Hadamard_permutation_problem}.

\begin{Algorithm}
\label{algrthm:Hadamard}
Initialize the joint system in the state $\ket{0}_c\, \ket{\Psi}_t$, with $\ket{\Psi}_t$ an arbitrary target state. Then, apply $H_{P}$ on $c$. Then, apply $S_N$ on the joint control-target system. Then, apply $H_{P}^{-1} (= H_{P}^{\mathsf T})$ on $c$. This gives the state
\begin{equation}
H_{P}^{-1}\,S_N\,H_{P} \ket{0}_c\, \ket{\Psi}_t= \ket{y}_c\, \Pi_0 \ket{\Psi}_t\ .
\label{eq:ACB_algorithm}
\end{equation}
Finally, read out $y$ as the outcome of a single-shot computational-basis measurement on $c$.
\end{Algorithm}

\begin{figure*}[t!]
\centering
\includegraphics[width=1.9\columnwidth]{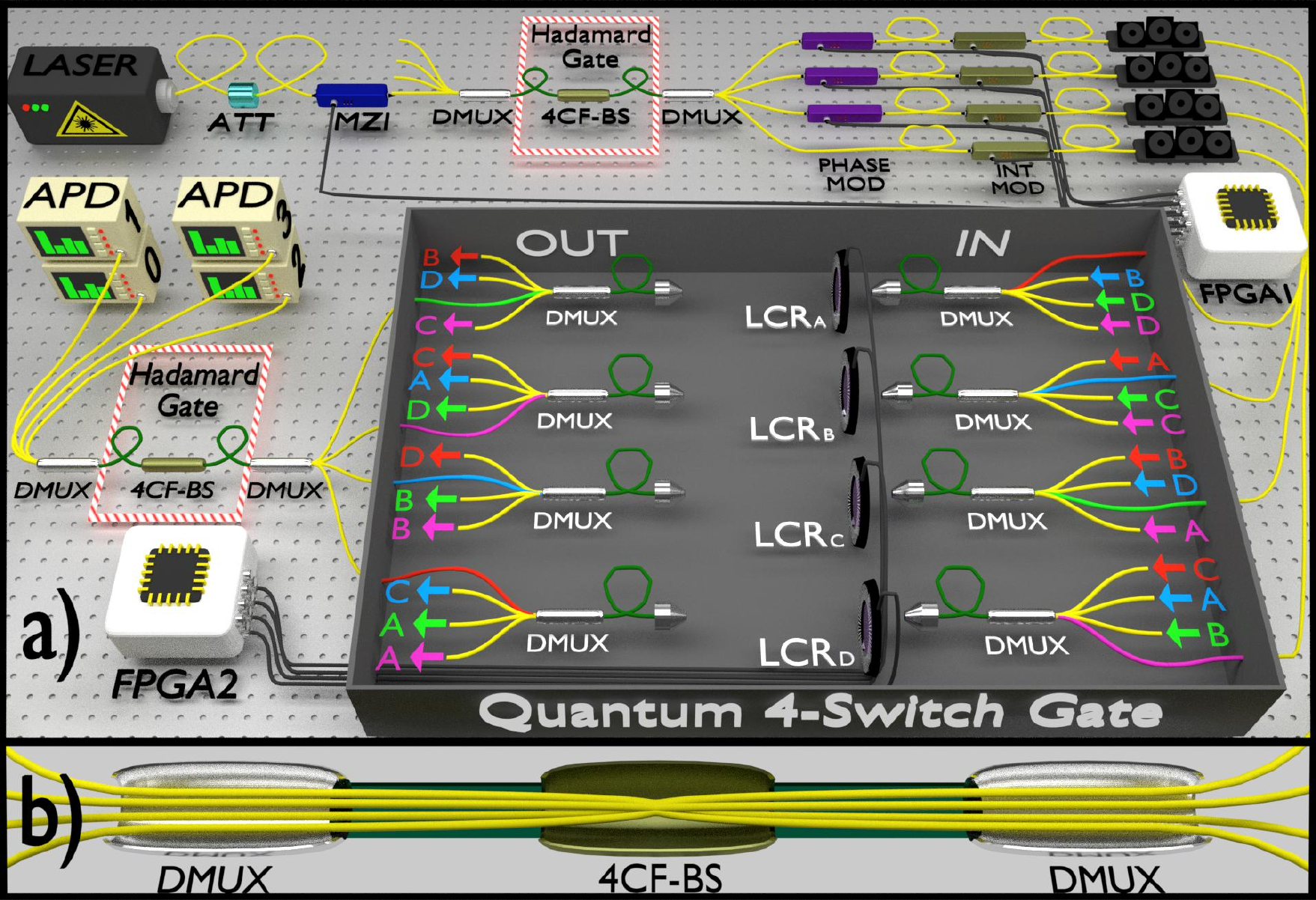}
\caption{ a) Illustration of our implementation of the quantum 4-switch gate ($S_4$). An input photon is divided coherently between four spatial modes using a four-core fiber beam splitter (4CF-BS), placed between commercial multiplexer/demultiplexer (DMUX) units, as shown in b). The four output modes are then sent to the quantum $4$-switch {$S_4$}. Each spatial mode is related to a unique permutation of the four unitary polarization operations applied by $S_4$ and indicated by a different color.  The photons enter through the $\mathsf{IN}$ side (right) and exit through the $\mathsf{OUT}$ side (left), where, for example, the notation ``$\leftarrow A$'' means ``from A'' and ``$ A \leftarrow$'' means ``to A''. One can follow a certain path by looking at the output labels. For instance, the {green} input mode enters in $C$ and continues to ``$B$, then $D$, then $A$, and finally exit'',  corresponding to the operation of the four polarization unitaries in the order {$CBDA$}.  After $S_4$, the four spatial modes are then recombined using a second 4CF-BS. Each output 0--3 is connected directly to a single-photon detector (APD). The detection of a single-photon in the $y$-th ($y=0,1,2,3$) output detector identifies in a single-shot the phase relation $y$ of the four unitaries implemented in the quantum $4$-switch gate. See the main text and Methods for further details.}
\label{fig:setup-concept}
\end{figure*}

This algorithm thus provides the desired phase relation between the $P$ different permutations of the $N$ unknown unitaries under consideration. The validity of Eq.~\eqref{eq:ACB_algorithm} is  proven explicitly in App.~\ref{subsec:proof_of_algorithm}. The query complexity of the algorithm is the same as that of the Ara\'ujo-Costa-Brukner algorithm: $Q=N$ for all $P\leq N!$.
The crucial resource for Algorithm \ref{algrthm:Hadamard} is the quantum $N$-switch process. 
Similarly to the Fourier promise problem \cite{Araujo2014}, no causally ordered process is known to solve Problem \ref{prblm:Hadamard_permutation_problem} in general (i.e., for any arbitrary set $\mathsf{U}$ of unknown gates fulfilling the promise) with a query complexity linear in $N$. In fact, the (query-wise) optimal causally ordered processes known to solve the problem in general are simply the fixed-gate circuits that simulate the quantum $N$-switch exactly (see Methods section), but these require considerably more queries\cite{Colnaghi2012, Araujo2014, Facchini2014}.
For instance, in the case where all gate permutations are considered ($P=N!$), simulating the quantum $N$-switch exactly in the blackbox scenario requires $Q=\Omega(N^2)$ oracle queries, i.e.\ quadratically higher in $N$. Another concrete example is the quantum 4-switch process for the $P=4$ permutations in the set $\{ABCD, BADC, CBDA, DACB\}$ [shown in Fig.~\ref{fig:4switchprocess} (b)], whose experimental implementation we describe below. The optimal circuit to simulate it exactly in the blackbox scenario requires $Q=9$ oracle queries, i.e.\ more than twice as many as with $S_4$ (see App.~\ref{subsec:simulation_N_switch}).

\section{Experimental quantum control of the order of multiple gate operations}
\label{sec:experiment}
The experiment is illustrated in Fig.~\ref{fig:setup-concept} (a). It is based on multi-core optical fibers and new related technology \cite{Richardson2013}, which was recently introduced as a toolbox for quantum information processing \cite{Canas2017,Ding2017,Xavier2020}. In our implementation of the quantum $4$-switch, the control system corresponds to the spatial mode of a single photon, while the target is its polarization. Following Algorithm \ref{algrthm:Hadamard}, a conventional illumination scheme (see Methods) is used to generate single photons propagating over a single-mode fiber in the initial spatial mode state $\ket{0}_c$. The photons are then sent through a four-core fiber beam splitter (4CF-BS), which has been shown to realize with high-fidelity the $H_{4}=\frac{M_4}{2}$ Hadamard operation given by \cite{Carine2020}
\begin{equation}
H_4=\frac12
\begin{bmatrix}
    1      & 1 & 1 & 1 \\
    1       & 1 & -1  & -1 \\
    1       & -1  & -1 & 1 \\
    1       & -1 & 1 & -1
\end{bmatrix}.
\label{eq:Had_mat}
\end{equation}
Note that this matrix is self-inverse. The 4CF-BS is placed between commercial spatial multiplexer/demultiplexer (DMUX) units \cite{Watanabe2012,Tottori2012}, which couple four single-mode fibers (yellow fibers) to the four cores of the multi-core fibers (green fibers). These units connect to the 4CF-BS through the multi-core fibers [see details in Fig.~\ref{fig:setup-concept}~(b)].

\renewcommand{\thetable}{\arabic{table}}
\begin{table}[t]
\begin{tabular}{|r|c|c|c|c|}\hline
\multicolumn{5}{|c|}{Table~1a}\\ \hline   
 $y$ 		&			0		 & 	1  &		 2		&		 3 		 \\ \hline 
 $U_A$	&$\mathbb1$& $Z$ &$\mathbb1$&		$Z$		\vphantom{$\frac{Z+X}{\sqrt{2}}$} \\ \hline 
 $U_B$	&		 $X$	 & $X$ & 		$X$ 	&		$X$		\vphantom{$\frac{Z+X}{\sqrt{2}}$} \\ \hline
 $U_C$	&$\mathbb1$& $Z$ & 		$Z$ 	&$\mathbb1$\\ \hline
 $U_D$	&		 $X$	 & $X$ &		$X$		&		$X$		 \\ \hline
\end{tabular}
\ \ \ \
\begin{tabular}{|r|c|c|c|c|}\hline
\multicolumn{5}{|c|}{Table~1b}\\ \hline
$y$ 	& 						0 				& 		 1 			&			2		 &		 3		\\ \hline
$U_A$	&$\frac{Z+X}{\sqrt{2}}$	& $\mathbb1$	&		 $Z$	 &		$Z$		\\ \hline
$U_B$	&$\frac{Z+X}{\sqrt{2}}$ & 		$X$			& 	 $X$	 &		$X$		\\ \hline
$U_C$	&				$\mathbb1$  		& 		$Z$			&$\mathbb1$&$\mathbb1$\\ \hline
$U_D$	&				$\mathbb1$			& $\mathbb1$	&$\mathbb1$&		$X$		\\ \hline
\end{tabular}
\caption{Tables of polarization unitaries used for the implementations of two different quantum 4-switch gates (both with the same set of gate permutations $\{ABCD, BADC, CBDA, DACB\}$; here $\mathbb1$ is the identity, $Z$ and $X$ are the Pauli operators). For both tables, each column provides a different set $\mathsf{U}$ of oracle gates. In turn, each such set exhibits the phase relations encoded -- via Eq.~\eqref{eq:property_Hadamard} -- in the corresponding column $y$ of the matrix in Eq.~\eqref{eq:Had_mat}. That is, the implemented  oracle gates fulfill the problem's promise with respect to the experimentally-implemented Hadamard matrix and the chosen set of permutations.}
\label{tab:4setsU}
\end{table}

After transmission through the 4CF-BS, the photon is sent to the quantum $4$-switch gate $S_4$, which will coherently apply different permutations of four unitary operations $U_i$ on the target system (photon polarization), depending on the spatial mode. To see this, note that each output of the 4CF-BS routes the photon through a different ordering of the polarization operations $U_i$, which are realized with controllable liquid crystal retarders (LCR). To control the implementation order of the $U_i$'s, we take advantage of the DMUX units. Each single-mode fiber input to the quantum $4$-switch gate is connected to a different four-core fiber on the $\mathsf{IN}$ side of $S_4$ using a DMUX unit. The other end of each 4CF is attached to a fiber launcher. The photon leaves the launcher in free space passing through the LCR and is coupled back into another 4CF on the $\mathsf{OUT}$ side. The $\mathsf{OUT}$ 4CF is connected (via another DMUX) to single mode fibers, which are then connected to the next 4CF (exploiting the already installed DMUXs) back on the $4$-switch's $\mathsf{IN}$ side, following the ordering showed in Fig.~\ref{fig:setup-concept}~(a). For example, a photon in the green input undergoes the operation of the four unitaries in the order {$C\rightarrow B\rightarrow D \rightarrow A$, resulting in the product unitary $\Pi_2 = U_A U_D U_B U_C$}. The other three inputs lead the photon through one of the other three permutations shown in the insets of Fig.~\ref{fig:4switchprocess} (b).  After $S_4$, a second Hadamard operation is applied to the control system using a second set of DMUX/4CF-BS/DMUX, in accordance with Algorithm~\ref{algrthm:Hadamard}. The setup is thus a four-arm interferometer with each output directly connected to an InGaAs single-photon detector (APD), working in gated mode and configured with 10$\%$ overall detection efficiency, and 5 ns gate width. The detection of a single-photon in the $y$-th ($y=0,1,2,3$) output detector univocally identifies in a single-shot the property  $y$ indicating the phase relations of the four unitaries implemented in the quantum $4$-switch gate.

Before implementing the quantum 4-switch, an initial alignment procedure using a polarimeter is performed. In-fiber polarization controllers (not shown in Fig.~\ref{fig:setup-concept}) are used in all single-mode fibers of the quantum $4$-switch to ensure that every fiber corresponds to an identity operation on the polarization. They are also used at the final set of DMUX/4CF-BS/DMUX to guarantee the indistinguishability of the core modes, such that there is no path-information available {that would} compromise the visibility of the interferometer \cite{Walborn2002,Torres-Ruiz2010}. The LCRs implementing the unitaries can be adjusted between identity and a half-wave plate by controlling the input voltage. In this way, we can toggle between an identity operation $\mathbb{1}$ and one of the Pauli operators $Z$, $(Z+X)/\sqrt{2}$ or $X$, when the orientation angle of the LCR is $0^\circ$, $22.5^\circ$ or  $45^\circ$, respectively. Importantly, we note that the LCRs were placed at the far-field plane of the 4CF launchers and that this guarantees that the unitary operations $U_i$ are indistinguishable when applied in different  orders (see Methods). A computer-controlled field programmable gate array (FPGA2) unit is used to control the LCRs. 

\begin{figure}[t]
\centering
\includegraphics[width=.75\columnwidth]{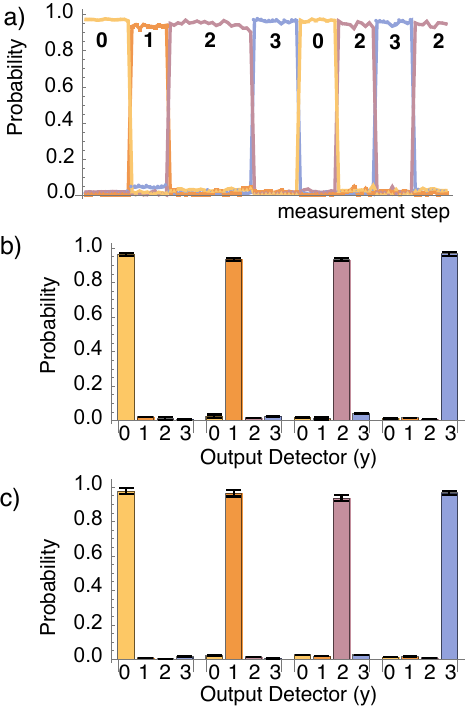}
\caption{a) A sequence of about 8 min of measurement results with our quantum $4$-switch process taken in real time. Measurements of 0.1s duration were taken continuously, realized within the phase stabilization routine (see Methods), in which the four sets of unitaries given each by the $y$-th column of Table~\ref{tab:4setsU}a were toggled randomly every minute. The number labels correspond to the columns of Table~\ref{tab:4setsU}a. Summary of experimentally obtained success probabilities to identify the commutation relations of the unitary operations in Table~\ref{tab:4setsU}a [panel b)] and Table~\ref{tab:4setsU}b [panel c)]. See text for more details.}
\label{fig:results_prob}
\end{figure}

In Table~\ref{tab:4setsU} we list the polarization operations $U_i$ for two different implementations of the quantum 4-switch. Table~\ref{tab:4setsU}a corresponds to orthogonal operations (for each given column), while Table~\ref{tab:4setsU}b includes non-orthogonal ones, which makes it more difficult to mimic the quantum $N$-switch with a causally ordered process (see below and App.~\ref{subsec:causal_attacks}). In each table, the \mbox{$y$-th} column defines a different set $\mathsf{U}$ of the target-system unitary gates and corresponds to the \mbox{$y$-th} column of the Hadamard matrix in Eq.~\eqref{eq:Had_mat} (see Methods). In our experiment, by exploiting the controlled LCRs, we are able to toggle between the different sets $\mathsf{U}$ of unitaries in real time. Fig.~\ref{fig:results_prob}~(a) shows an example of the results recorded while switching randomly (with uniform probabilities) between operations corresponding to different columns of Table~\ref{tab:4setsU}a, about every minute. In each 0.1~s measurement we detected a total of $\sim 6000$ events.  Figs.~\ref{fig:results_prob}~(b) and~(c) show a summary of experimentally obtained success probabilities (each obtained from $\sim 3\times10^4$ events) to identify the relative-phase relations between the different permutations of the unitary operations in Table~\ref{tab:4setsU}a and Table~\ref{tab:4setsU}b, resp. For Table~\ref{tab:4setsU}a we obtain an average success probability of $p_\text{succ}=0.948 \pm 0.005$, whereas for Table~\ref{tab:4setsU}b we obtain $p_\text{succ}=0.959 \pm 0.008$. Error bars correspond to one standard deviation, and are obtained by error propagation of the Poissonian count statistics.   These results demonstrate the successful implementation of the quantum 4-switch process.

\section{Benchmarking experimental quantum control of multiple gate orders}
\label{sec:witness_discussion}
To benchmark the realization of QCGO, it is useful to imagine a verification scenario, in which a Verifier controls the oracle, while the process is implemented by a Prover. 
The Prover wishes to prove to the Verifier that the process does display QCGO, and the Verifier can test this by asking the Prover to compute properties of oracles involving different gates. 
The quantum $N$-switch process allows the Prover to solve the computations with considerably fewer oracle queries than any process with fixed (or classically controlled) gate connections. Indeed, it is the only process known to provide a unit success probability for Problem~\ref{prblm:Hadamard_permutation_problem} in general (i.e.\ for any set of black-box gates satisfying the promise) with only $N$ queries to the oracle. This can be used to give the Verifier evidence in favour of the Prover's honesty.
However, if the table of oracle-gates has a small number of columns -- e.g., as in Table~\ref{tab:4setsU} -- a dishonest Prover with side information about the table can attain $p_\text{succ}=1$ with a causally ordered process (see App.~\ref{subsec:causal_attacks}), thus deceiving the Verifier. 

One way to benchmark experimental quantum switches with minimal assumptions  is by measuring so-called causal witnesses \cite{Araujo2015,Branciard2016a}. Interestingly, by increasing the number of columns in the oracle-gate table (i.e., of possible choices for the gate sets $\mathsf{U}$) and suitably choosing their prior probability distribution, Algorithm \ref{algrthm:Hadamard} can be turned into a causal witness for the quantum switch. That is, for sufficiently large oracle-gate tables and an appropriate prior distribution the gate sets $\mathsf{U}$, an upper bound $p_\text{succ}^\text{CCGO}$ strictly smaller than one can be found for the probability of success attainable by processes with \emph{classical} control of gate orders. This provides us with a gap from the the probability of success obtained by the quantum switch, which always remains unity in the noiseless case. Details on our search for witnesses are given in App.~\ref{subsec:witness}.

Unfortunately, the number of measurement settings required to measure such witnesses is prohibitively high in practice for this experimental setup. For instance, the best witness for $W_4$ we could obtain with the above-mentioned approach gives $p_\text{succ}^\text{CCGO}\approx 0.89$, but requires an oracle-gate table with 300 columns. Alternatively, weaker witnesses with $p_\text{succ}^\text{CCGO}\approx 0.92$ can also be found, but these still require 60 columns. Our LCR-based setup cannot switch among so many gates in a practical way. Nevertheless, it is yet a remarkable feature of our experiment that we do reach values of $p_\text{succ}$ significantly higher than both bounds, which would conclusively benchmark $W_4$ for higher number of settings. In addition, we note that witnesses with similarly high numbers of settings (259) have indeed been measured in other platforms, though with much slower switching times \cite{Rubino2016}.

Alternatively, smaller oracle-gate tables suffice if the Verifier can actively reduce the Prover's potential knowledge about the tables. One way to do this is by allowing the Verifier to apply a random basis rotation to each gate before delivering it to the Prover.
For instance, in this scenario, an upper bound $p_\text{succ}^\text{CCGO}\approx 0.84$ can be obtained for an oracle-gate table with only 30 columns (see App.~\ref{subsec:witness}).
Unfortunately, implementing such a causal witness would require the ability to switch among a continuum of gates, which is again experimentally infeasible. 
Nevertheless, here we are mainly interested in benchmarking our implementation of $W_4$ against experimental imperfections, rather than against hypothetical malicious Provers exploiting side-information about the gates' bases. 
In this regard, the experimentally obtained values in Fig. \ref{fig:results_prob} are in the range $p_\text{succ} \approx 0.93$-$0.97$, which suggests that our setup should be capable of obtaining average success probabilities that are larger than the thresholds mentioned above, for a larger number of settings.  Though not yet conclusive, this provides encouraging evidence for the QCGO of the implemented process.

\section{Discussion}
\label{sec:conclusions}

Here we introduced the ``Hadamard promise problem'', a novel computational primitive involving the relative phases between different permutations of multiple unknown gates. We presented an algorithm to solve it  efficiently,  illustrating a quantum computational advantage associated to the coherent quantum control of the order in which a sequence of $N$ unitary operations is applied. Our algorithm, which we implemented experimentally for $N=4$, exploits the quantum $N$-switch process to solve the problem with $N$ applications of the unitary gates, whereas the known methods exploiting fixed gate orders use the gates $O(N^2)$ times. Both problem and algorithm have the advantage that the target system needs only be two-dimensional, as opposed to $N!$-dimensional as in previous proposals. This could inspire new approaches for exploiting indefinite causal order in quantum computation and communication, as well as for studying causal non-separability in physical systems.

We experimentally implemented the algorithm by constructing a quantum 4-switch process that coherently controls four different gate orderings with high fidelity, showing success probabilities for the algorithm of $\sim 0.95$.  The all-optical setup involves a four-path interferometer constructed with new multi-core optical fiber technology. As discussed in the Methods, the best known quantum circuit with fixed gate orders solves this problem with $9$ gate queries. Our experiment thus corresponds to a 5-query improvement. Moreover, this is, to the best of our knowledge, the first report of a quantum superposition of more than 2 temporal orders. In addition, our implementation presents some technical advantages as well: On the one hand, it is versatile in that the gate orders can be modified in a practical fashion by switching the optical fiber connections and that the unitary gates themselves can be automatically controlled through  the liquid crystal polarization retarders. On the other hand, the setup can be scaled up to higher control-system dimensions in a straightforward fashion. This work constitutes a key step towards realizing and verifying causal non-separability among a large number of parties, and should play an important role in developing methods to exploit this resource.

\section*{Methods}
\subsection{Query complexity analysis}
\label{sec:query_comp}
One may argue that implementing $S_N$ is not the only way to solve Problem \ref{prblm:Hadamard_permutation_problem} (which is also true for the Fourier promise problem \cite{Araujo2014}). Here, we estimate the query complexity of other plausible approaches.

A natural approach one may attempt is to tomographically reconstruct the $N$ unitary gates and then multiply them to estimate the $\Pi_x$'s, from which one can infer $y$. Since each $\Pi_x$ is an $N$-fold product of the $U_i$'s, the overall error $\varepsilon$ in its estimation is $\varepsilon=\Omega\left(N\, \epsilon \right)$, where $\epsilon$ is the statistical error of the reconstruction of each $U_i$. To attain a constant overall error one thus needs $\epsilon=O\left(1 / N\right)$, which, by virtue of Hoeffding's bound, in turn requires $q=O\left(1/\epsilon^2\right)=O\left(N^2\right)$ queries to each $U_i$. Moreover, since there are $N$ gates to reconstruct, the overall query complexity is $Q=O\left(N\, q\right)=O\left(N^3\right)$, i.e.\ cubically worse in $N$ than with the quantum $N$-switch.
Another alternative is to tomographically reconstruct each $\Pi_x$ directly,
and from that infer $y$. 
However, to query each $N$-fold product $\Pi_x$ one must query all $N$ unitaries; and there are $P$ such products. Hence, the overall query complexity is $Q=O\left(N\, \, P\right)\geq O\left(N^2\right)$ if one considers $P\geq N$ (as we did in our experimental demonstration), i.e.\ quadratically worse in $N$ than with the quantum $N$-switch. A third possibility could be to directly estimate the signs of the commutators between the $\Pi_x$'s,
and from that infer $y$. A canonical tool for that is the well-known Hadamard test \cite{Hadamard}. This allows one to estimate overlaps of the form $\bra{\Psi}_t\Pi_x\ket{\Psi}_t$ or $\bra{\Psi}_t\Pi^{\dagger}_x\,\Pi_{x'}\,\Pi_x\ket{\Psi}_t$ directly from  queries to $\Pi_x$ or $\Pi_x$ and $\Pi_{x'}$, respectively, for any state $\ket{\Psi}_t$. As before, each query to $\Pi_x$ accounts for $N$ queries to the gates, and  the overall query complexity is again $Q=O\left(N\, \, P\right)\geq O\left(N^2\right)$. 

Finally, one can simulate $S_N$ exactly with a circuit with fixed gate orders. For the usual case where all $P=N!$ permutations are considered, the optimal causally ordered circuit that synthesizes $S_N$ in the blackbox scenario displays complexity $Q=\Omega(N^2)$ \cite{Colnaghi2012, Araujo2014, Facchini2014}. For the concrete case experimentally studied here, $P=N=4$, the optimal causally ordered circuit that synthesizes $S_4$ requires 9 queries  (see App.~\ref{subsec:simulation_N_switch}). In fact, this is the reason why we chose the particular permutation set $\{ABCD, BADC, CBDA, DACB\}$. Through a brute-force search, we found that, from all quartets of permutations, most of them require 7 queries or less  with the simulation strategy presented in App.~\ref{subsec:simulation_N_switch}, some other 8 queries, and a few of them (including the one chosen here) require the maximum of 9 queries. Thus, the specific version of the quantum 4-switch implemented here provides a gap of $9-4=5$ queries with respect to  all causally ordered processes.

\subsection{Experimental details}
{\em Single photon source.---} The single-photon light source is composed of a semiconductor distributed feedback telecom laser ($\lambda = 1546$ nm) connected to an external fiber-pigtailed amplitude modulator (MZI). An FPGA unit (FPGA1) was used with the MZI to externally modulate the laser and generate optical pulses 5 ns wide. Optical attenuators (ATT) are used before MZI to  create weak coherent states with a mean photon number per pulse of $\mu=0.2$. In this case,  $90\%$ of the non-null pulses generated contain a single photon. Thus, our source is a good approximation to a non-deterministic single-photon source, which is commonly adopted in quantum communications \cite{Gisin2002}. FPGA1 also controls the active phase stabilization of the system and registration of single-photon counts at each of the four detectors during the measurement procedure (see below).

\begin{figure}
\centering
\includegraphics[width=0.75\columnwidth]{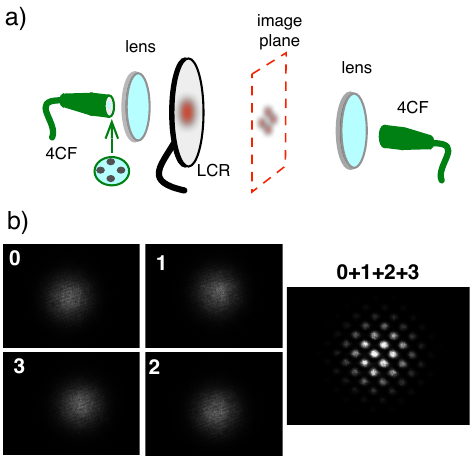}
\caption{a) Illustration of the 4CF launchers and the liquid crystal retarders (LCR) implementing the unitaries $U_i$. The LCRs are placed at the Fourier plane of the output coupling lenses.  b) Images recorded at the LCRs plane, of each core alone, as well as the output when all cores are connected, showing large spatial overlap between the cores modes. This guarantees that $U_i$ 's are indistinguishable when applied in different  orders.}
\label{fig:setup-details}
\end{figure}

{\em Indistinguishability of the multi-gate operations in different orders.---} The four unitary operators $U_i$ ($i=A,B,C,D$) were realized using birefringent liquid crystal retarders. An important aspect of the experiment is to guarantee the realization of the same unitary operation $U_i$, for all different orders considered. That is, the implementation of $U_i$ must be independent of the illuminated core on the corresponding 4CF at the $\mathsf{IN}$ side of the oracle. To achieve this, the LCRs are placed in the Fourier plane of the objective lenses of the 4CF fiber launchers [see Fig.~\ref{fig:setup-details} (a)]. At the exit face of this fiber, the output single mode of each core is given by a gaussian function $g(\vec{r})$ centered at the core position $\vec{r}_c$. At the Fourier plane of the launcher lens, the spatial distribution of each core is given by the Fourier transform $\mathcal{F}[g(\vec{r}-\vec{r}_c)](\vec{s}) \propto \exp(i k\vec{s}\cdot \vec{r}_c/f) g(\vec{s})$. Therefore, irrespective of the illuminated core, all core modes overlap at the same central point with the intensity proportional to $|g(\vec{s})|^2$. 
This avoids spatial distinctions as in certain implementations for $N=2$ gates \cite{Procopio2014,Rubino2016}. To guarantee this condition for our experiment, we used a CCD camera to record the intensity distributions at the Fourier plane (with the LCRs removed), as shown in Fig.~\ref{fig:setup-details}~(b). The images, obtained with an intense laser, show the centering of {the} light distribution when a single core is connected. The resulting interference pattern when all cores are illuminated shows high-visibility, confirming spatial indistinguishability. This guarantees that the unitary operations $U_i$ are indistinguishable when applied in different orders-- a crucial requirement for a valid implementation of an $N$-switch \cite{Goswami2018}.

{\em Phase stabilization and Measurement procedure.---} Phase (PHASE MOD) and intensity modulators (INT MOD) are used after the first 4CF-BS, on each arm of the interferometer (see Fig.~\ref{fig:setup-concept} (a)), to set the relative phases between the four spatial modes to zero, and to adjust the amplitudes. The FPGA1 unit is used to implement a control system to actively compensate phase-drifts in the quantum 4-switch. The control is based on a perturb and observe power point tracking method \cite{Bhatnagar2013,Carine2020}. Basically, the phase drift compensation algorithm will perturb the $k$th phase modulator to cancel any phase noise using a high-speed signal. The algorithm does this sequentially to each phase modulator and in each step it maximizes the number of photo-counts in the output detector ``0'' with the LCRs set to realize column $y=0$ of one of the tables in Table~\ref{tab:4setsU}. When the counts achieve a given threshold value for the success probability, the voltages applied to the phase modulators are maintained constant, and an $\mathsf{ON}$ signal is sent to FPGA2  to activate the LCRs by applying a constant voltage, realizing any one of the four columns of the respective table in Table~\ref{tab:4setsU}, chosen by the user. After a {0.2~s} deadtime to allow for the LCRs voltages to reach the desired value, a {0.1~s} measurement stage is realized. After a single measurement window, an $\mathsf{OFF}$ signal is sent to return the LCRs to column 0. In this way, we can switch rapidly between columns 0-3 of the tables. The control system monitors the phase stabilization of the interferometer in real-time after every measurement.  
\par
We have used this phase stabilization routine in other work \cite{Carine2020}, and obtained visibilities over 99\%.  Here, our success probability is limited to about 95\% due to slightly imperfect polarization rotations of the LCRs, as well as the difficulty in achieving proper alignment of the polarization state for the different LCR combinations in each path, which we observed in the initial alignment procedure using the polarimeter (see section IV).

\section*{Acknowledgements}
We thank Barbara Amaral, Johanna Barra, Fabio Costa and {\v C}aslav Brukner for helpful insights. MMT and LA acknowledge financial support from the Brazilian agencies CNPq (PQ grant No. 311416/2015-2 and INCT-IQ), FAPERJ (PDR10 E-26/202.802/2016 and JCN E-26/202.701/2018), CAPES (PROCAD2013), and the Serrapilheira Institute (grant number Serra-1709-17173). This work was also supported by Fondo Nacional de Desarrollo Cient\'{i}fico y Tecnol\'{o}gico (ANID) (3200779, 1190901, 1200266, 1200859) and  ANID – Millennium Science Initiative Program – ICN17\_012.  JC was supported by ANID/REC/PAI77190088. AA was supported by the Swiss National Science Foundation (Starting Grant DIAQ and NCCR SwissMAP). MA has received funding from the European Union’s Horizon 2020 research and innovation programme under the Marie Skłodowska‐Curie grant agreement No 801110 and the Austrian Federal Ministry of Education, Science and Research (BMBWF). It reflects only the author's view, the EU Agency is not responsible for any use that may be made of the information it contains.

%



\section*{Appendix}
\appendix
\label{App}

\subsection{Proof of Eq.~\eqref{eq:ACB_algorithm}.}
\label{subsec:proof_of_algorithm}
First, note that (just like the Fourier transform) the Hadamard gate $H_P$ maps $\ket{0}_c$ to the uniform superposition of all computational-basis states (under the assumption that the corresponding Hadamard matrix $M_P$ only has $+1$ values along the first column):
\begin{align}
H_{P} \ket{0}_c\, \ket{\Psi}_t = &\frac{1}{\sqrt{P}}\sum_{x\in[P]}\ket{x}_c\ket\Psi_t.
\label{eq:appliedHadamard}
\end{align}
Then, the quantum $N$-switch gate introduces the sign  $m_{x,y}$ to each computational-basis state $\ket{x}_c$ in the superposition:
\begin{align}
\nonumber
S_N\,H_{P} \ket{0}_c\, \ket{\Psi}_t = &\frac{1}{\sqrt{P}}\sum_{x\in[P]}\ket{x}_c \Pi_x \ket\Psi_t\\
= &\left(\frac{1}{\sqrt{P}}\sum_{x\in[P]}m_{x,y}\ket{x}_c \right)\Pi_0 \ket\Psi_t,
\label{eq:appliedSwitch}
\end{align}
where the second equality follows from Eq.~\eqref{eq:property_Hadamard}. Now, by definition, the state within the brackets is $H_{P} \ket{y}_c$. Hence, applying $H_{P}^{-1}$ to both sides of Eq.~\eqref{eq:appliedSwitch} yields Eq.~\eqref{eq:ACB_algorithm}. \hfill$\square$ 

\subsection{Exact simulation of the quantum $N$-switch with a fixed-gate-order circuit}
\label{subsec:simulation_N_switch}
\begin{figure*}
\includegraphics[width=.99\textwidth]{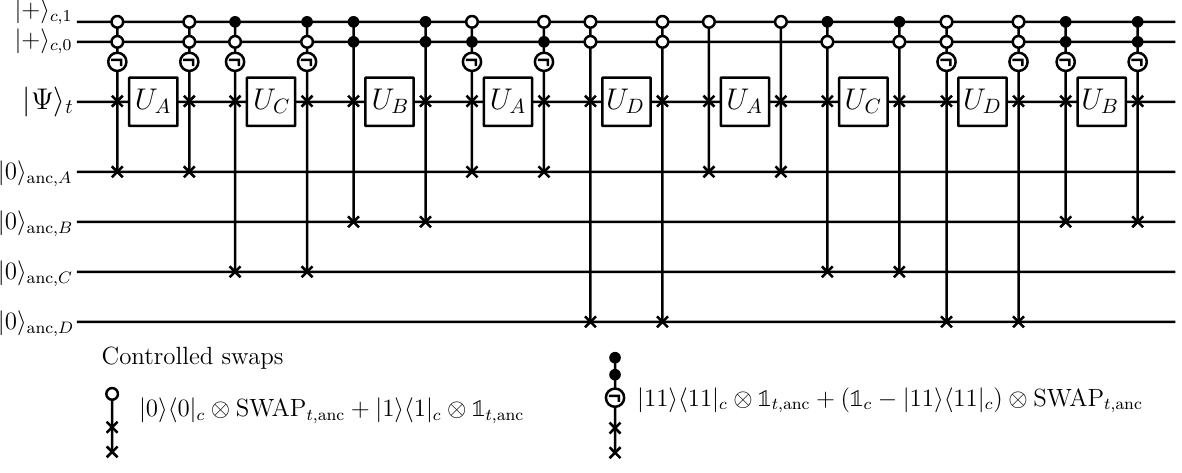}
\caption{Fixed-gate-order circuit that simulates the quantum $4$-switch that was realized experimentally, i.e.\ with quantum control of the four gate sequences $\Pi_0=U_DU_CU_BU_A$, $\Pi_1=U_CU_DU_AU_B$, $\Pi_2=U_AU_DU_BU_C$, and $\Pi_3=U_BU_CU_AU_D$. Before and after each unitary $U_i$, a pair of controlled swap gates controls whether $U_i$ is applied to the target system or to an ancilla; the control qudit has dimension $P=4$, here represented as two qubits (with $x = 0, 1, 2$ and $3$ encoded as $00, 01, 10$ and $11$, resp.). Black dots indicate an operation conditioned on the $\ket1_c$ state, white dots, conditioned on the $\ket0_c$ state. Conditioning on negation of certain states is also needed, as exemplified in the legend below the circuit.}
\label{fig:circuitsimulation}
\end{figure*}

It is possible to simulate the quantum $N$-switch -- i.e.\ produce the same superposition of unitaries $\{\Pi_x\}_{x\in[P]}$ as the quantum $N$-switch for whatever unitaries $U_i$ are inserted at its open slots -- with a causally ordered circuit at the cost of making more uses (queries) of each unitary. The basic idea behind such circuit is to apply the unitaries coherently controlled by a qudit. However, this is not a straightforward task with black-box unitaries \cite{Araujo2014a,Thompson2013,Friis2014,Abbott2020,Chiribella2019,Kristjansson2020}. A workaround is to use ancillas and controlled swap gates that coherently control whether each target-system gate is effectively applied to the target system or to an ancilla. This can be done with a circuit such as in Fig.~\ref{fig:circuitsimulation}, which uses a $P$-dimensional control qudit and $N$ $d$-dimensional ancilla systems (one for each gate $U_i$).
Importantly, as the reader may verify, all $N$ ancilas experience the same overall gate sequence for all input states of the control register, which guarantees that the ancillas disentangle from the target and control systems by the end of the circuit. For instance, for the circuit in Fig.~\ref{fig:circuitsimulation}, the final state of the ancillas is $U_A^2\ket0_{\text{anc},A}\ U_B\ket0_{\text{anc},B}\ U_C\ket0_{\text{anc},C}\ U_D\ket0_{\text{anc},D}$.

With this circuit scheme, the problem of simulating the superposition of unitaries produced by a quantum $N$-switch reduces to finding a supersequence that includes all the desired permutations as subsequences; the query complexity of this scheme is then given by the length of the shortest such supersequence \cite{Koutas1975,Facchini2014}. In the experiment and Fig.~\ref{fig:circuitsimulation}, $ACBADACDB$ is the supersequence to the quartet of permutations $\{ABCD,BADC,CBDA,DACB\}$ (notice that the subsequences need not be contiguous). We have made an extensive numerical search of all quartets of permutations of $A$, $B$, $C$, $D$. There are $\binom{N!-1}{P-1} = \binom{23}{3} =$ 1771 unique quartets, where quartets that differ only by relabeling are disconsidered (this amounts to, for instance, only considering quartets that include some fixed permutation, e.g.\ $ABCD$). Of those, most require a supersequence of length 8 or less (37 unique quartets require length 6; 946 require length 7; 779 require length 8) and only 9 require length 9. Since the higher the supersequence length, the higher the query complexity of the simulation by fixed-gate-order circuit, we chose one of the latter 9 quartets for our experiment (as well as Fig.~\ref{fig:circuitsimulation}).
 Notice that all 9 black boxes are queried once, irrespective of whether they are effectively used in the superposition or not, hence the query complexity of this simulation of the quantum $4$-switch is 9.

\subsection{Fixed-gate circuit algorithms for the Hadamard promise problem exploiting side information about the gates}
\label{subsec:causal_attacks}

Let us revisit the adversarial scenario of a Verifier who controls the oracle and poses the Hadamard promise problem to a Prover. The Prover thus receives unknown (to them) unitaries and uses them to the best of their abilities to solve the problem and output the correct answer to the Verifier.
As we showed, a Prover in possession of a quantum $N$-switch can solve the problem with 100\% success rate using only a single query from each unitary. We now ask: can a Prover solve the problem with access only to fixed-gate-order circuits?

By performing the simulations in the previous section, they are also able to solve the Hadamard promise problem with 100\% success rate. However, they must request additional queries of the oracle to the Verifier, a tell-tale sign to the latter that the quantum $N$-switch has not been realized.

We now explore the case of a Prover with side information on the unitaries from the oracle. More specifically, let us suppose they know the table of unitaries that the Verifier uses (Table~\ref{tab:4setsU}a or \ref{tab:4setsU}b), but not which column is selected in each run. This information aids the Prover, who may no longer need to produce the superposition of unitaries from the previous section.

If Table~\ref{tab:4setsU}a is used, the Prover's strategy is relatively simple. By inputting a $\ket+:=(\ket0+\ket1)/\sqrt2$ state to black box $U_A$, the output state will be either $\ket+$, if $U_A=\mathbb1$, or $\ket-:=(\ket0-\ket1)/\sqrt2$, if $U_A=Z$. With a measurement of the output in the $X$ basis, they can identify $U_A$ (we call this an $X$-basis test on $U_A$). Doing the same procedure on $U_C$, they identify this unitary as well and discover the column $y$ of Table~\ref{tab:4setsU}a being used. Since only 1 query or less of each unitary is needed, the Prover can in fact deceive the Verifier in this case.

If instead Table~\ref{tab:4setsU}b  is used, the Prover requires a slightly more complex fixed-gate-order circuit to deceive the Verifier. It begins with an $X$-basis test on applied to $U_C$, which reveals the content of that black box. In turn, $U_D$ is revealed with an analogous $Z$-basis test, with input state $\ket0$ and measurement of output in the $Z$ basis. If one of these two black boxes is revealed to be a Pauli operator ($Z$ or $X$, resp.), then that run of the promise problem has been solved ($y=1$ or $3$, resp). However, if both $U_C=\mathbb1$ and $U_D=\mathbb1$, both $y=0$ and $y=2$ are possible, and the black boxes $U_A$, $U_B$ need to be used. Since the quantum $N$-switch finds the correct value of $y$ with probability one, so is the goal of the Prover here. However, the two possible unitaries for $U_A$ ($\frac{Z+X}{\sqrt{2}}$, $Z$) are not orthogonal, i.e.\ not perfectly distinguishable, and the same happens with $U_B$. No independent use of $U_A$ and $U_B$ can tell the columns apart with certainty. There is a viable strategy, though, using $U_A$ and $U_B$ in sequence. Notice indeed that $U_BU_A=\mathbb1$ for column 0 and $U_BU_A=-iY$ for column 2. A $Z$- or $X$-basis test applied to the sequence of the two unitaries $U_A$ and $U_B$ can distinguish these two possibilities, again solving the problem with certainty.

If the Prover does not know whether the Verifier uses Table~\ref{tab:4setsU}a or \ref{tab:4setsU}b, the former needs to first identify which table is used.  This table identification can be done with a $Z$-basis test on $U_D$, which reveals whether $U_D=X$ or $U_D=\mathbb1$. The strategy for Table~\ref{tab:4setsU}a is applied in the former case, that for Table~\ref{tab:4setsU}b in the latter (notice that column $y=3$ is the same for both tables).

\subsection{Causal witnesses for the 4-switch process}
\label{subsec:witness}

In order to certify, via the Hadamard promise problem, that a given process exhibits some quantum control of gate orders (QCGO), one may look for the maximal probability of success $p_\text{succ}^\text{CCGO}$ that processes with \emph{classical} control of gate orders (CCGO) can reach: If this upper bound is strictly smaller than $1$, it becomes possible to experimentally obtain a probability of success $p_\text{succ} > p_\text{succ}^\text{CCGO}$ and thus prove that these results cannot be explained by CCGO.

For a fixed choice of gate permutations and of Hadamard matrix under consideration, the ``causal bound'' $p_\text{succ}^\text{CCGO}$ still depends on the specific choice of possible sets $\mathsf U$, and of the prior distribution with which each set is chosen in each experimental run.
Considering different possible sets $\mathsf U_k$, each satisfying the promise of Eq.~\eqref{eq:property_Hadamard} for some value $y = y_k$ and chosen with probability $q_k$, the probability of success (i.e., of obtaining the correct value $y = y_k$) of the Hadamard promise problem is obtained as
\begin{equation}
p_\text{succ} = \sum_k \, q_k \, \text{Prob}(y = y_k|\mathsf U = \mathsf U_k). \label{eq:Psucc}
\end{equation}

To compute the above probabilities, and to obtain the causal bound $p_\text{succ}^\text{CCGO}$, we use the so-called ``process matrix framework'' \cite{Oreshkov2012}.
In this framework the process under consideration (i.e., in our case, the circuit that connects the 4 unitaries and the final measurement) is described by the ``process matrix'' $W$, acting on the tensor product of all input and output Hilbert spaces of the 4 unitaries and of the final measurement. When the 4 qubit unitaries from some quartet $\mathsf U_k = \{U_A^{(k)},U_B^{(k)},U_C^{(k)},U_D^{(k)}\}$ are applied, the probability $\text{Prob}(y = y_k|\mathsf U = \mathsf U_k)$ that the final measurement in the computational basis $\{\ket{y}_c\}_{y\in[4]}$ of $\mathbb H_c$ gives the outcome $y_k$ for an arbitrary process matrix $W$ is obtained as
\begin{equation}
\text{Prob}(y = y_k|\mathsf U = \mathsf U_k) = \Tr \big[ \big( \KetBra{\mathsf U_k}^T \!\otimes \ketbra{y_k}{y_k}_c \big) W \big]
\end{equation}
with
\begin{equation}
\Ket{\mathsf U_k} \coloneqq \Ket{U_A^{(k)}} \Ket{U_B^{(k)}} \Ket{U_C^{(k)}} \Ket{U_D^{(k)}}.
\end{equation}
Here, $^T$ denotes the transposition in the computational basis $\{\ket{0},\ket{1}\}$ of $\mathbb H_t$ and $\Ket{U_i^{(k)}}\in \mathbb H_t\otimes \mathbb H_t$ is the Choi vector representation \cite{Choi1975} of the $i$th unitary $U_i^{(k)}$, for $i = A, B, C$ or $D$, technically defined as $\Ket{U_i^{(k)}}\coloneqq \mathbb1 \otimes U_i^{(k)} \Ket{\mathbb1}$, with  $\Ket{\mathbb1} \coloneqq \ket{0}\!\ket{0} + \ket{1}\!\ket{1}$. 
According to Eq.~\eqref{eq:Psucc}, the success probability is then obtained as
\begin{align}
& p_\text{succ} = \Tr \big[ G \, W \big] \notag \\
& \quad \text{with} \quad G = \sum_k \, q_k \, \KetBra{\mathsf U_k}^T \!\otimes \ketbra{y_k}{y_k}_c. \label{eq:Psucc_Tr_GW}
\end{align}

The process matrix describing the ideal 4-switch process of Fig.~\ref{fig:4switchprocess} (b) is given by \cite{Araujo2015,Oreshkov2016} $W_4 = \ketbra{w_4}{w_4}$, where
\begin{multline}
\label{eq:wswitchket}
\ket{w_4} = \ket{0}^{c_p}\Ket{\id}^{t_pA_I}\Ket{\id}^{\!A_OB_I}\Ket{\id}^{\!B_OC_I}\Ket{\id}^{\!C_OD_I}\Ket{\id}^{\!D_Ot_{\!f}}\ket{0}^{c_{\!f}} 
\\+ \ket{1}^{c_p}\Ket{\id}^{t_pB_I}\Ket{\id}^{\!B_OA_I}\Ket{\id}^{\!A_OD_I}\Ket{\id}^{\!D_OC_I}\Ket{\id}^{\!C_Ot_{\!f}}\ket{1}^{c_{\!f}}
\\+ \ket{2}^{c_p}\Ket{\id}^{t_pC_I}\Ket{\id}^{\!C_OB_I}\Ket{\id}^{\!B_OD_I}\Ket{\id}^{\!D_OA_I}\Ket{\id}^{\!A_Ot_{\!f}}\ket{2}^{c_{\!f}}
\\+ \ket{3}^{c_p}\Ket{\id}^{t_pD_I}\Ket{\id}^{\!D_OA_I}\Ket{\id}^{\!A_OC_I}\Ket{\id}^{\!C_OB_I}\Ket{\id}^{\!B_Ot_{\!f}}\ket{3}^{c_{\!f}} 
\end{multline}
and the superscripts indicate the Hilbert spaces in which the various states are defined: $c_p,c_f$ refer to the past and the future of the control system, $t_p,t_f$ refer to the past and the future of the target system, $A_I$ and $A_O$ refer to the input and output spaces of operation $U_A$, and similarly for the other parties. Notice that, for the sake of clarity, Fig.~\ref{fig:4switchprocess}a) uses a simplified notation based on the necessary isomorphism between $t_p$, $t_f$, $A_I$, $A_O$, and the other parties' inputs and outputs (as well as between $c_p$ and $c_f$).

In Algorithm~\ref{algrthm:Hadamard} we input the initial control state $H_P\ket{0}$ into $c_p$, the initial target state $\ket{\Psi}$ into $t_p$, and apply $H_P^{-1}$ to the resulting state of the control system in $c_{\!f}$. These fixed steps can be incorporated into the process-matrix description. The resulting matrix that describes our effective process is then $W_4' = \Tr_{t_{\!f}} \ketbra{w_4'}{w_4'}$ with
\begin{align}
\ket{w_4'} = & \, \frac12 \Big[ \ket{\Psi}^{\!A_I}\!\Ket{\id}^{\!A_OB_I}\Ket{\id}^{\!B_OC_I}\Ket{\id}^{\!C_OD_I}\Ket{\id}^{\!D_Ot_{\!f}} H_P^{-1}\! \ket{0}_c  \notag
\\ & \ +\ket{\Psi}^{\!B_I}\!\Ket{\id}^{\!B_OA_I}\Ket{\id}^{\!A_OD_I}\Ket{\id}^{D_OC_I}\Ket{\id}^{C_Ot_{\!f}} H_P^{-1}\! \ket{1}_c \notag
\\ & \, + \ket{\Psi}^{\!C_I}\!\Ket{\id}^{\!C_OB_I}\Ket{\id}^{\!B_OD_I}\Ket{\id}^{\!D_OA_I}\Ket{\id}^{\!A_Ot_{\!f}} H_P^{-1}\! \ket{2}_c \notag
\\ & + \ket{\Psi}^{\!D_I}\!\Ket{\id}^{\!D_OA_I}\Ket{\id}^{\!A_OC_I}\Ket{\id}^{\!C_OB_I}\Ket{\id}^{\!B_Ot_{\!f}} H_P^{-1}\! \ket{3}_c \! \Big].
\end{align}
Using this process matrix we can verify that for any set $\mathsf U_k = \{U_A^{(k)},U_B^{(k)},U_C^{(k)},U_D^{(k)}\}$ satisfying the promise~\eqref{eq:property_Hadamard} for some $y = y_k$, one has $\Tr \big[ \big( \KetBra{\mathsf U_k}^T \!\otimes \ketbra{y_k}{y_k}_c \big) W_4' \big] = 1$, so that the success probability of  Algorithm~\ref{algrthm:Hadamard}, using the 4-switch, is indeed unity.

Processes that display CCGO, on the other hand, are described by process matrices from a particular subset of all possible matrices, with some specific structure.
In Ref.~\citenum{Wechs2020}, it is indeed shown that (in our scenario, with 4 operations and a final measurement) CCGO process matrices $W$ must have a decomposition of the form
\begin{align}
W = \sum_{(i,j,k,l)} W_{(i,j,k,l),c} \label{eq:CCGO_decomp}
\end{align}
in terms of positive semidefinite matrices (not necessarily valid process matrices) $W_{(i,j,k,l),c}$, for all $4! = 24$ permutations $(i,j,k,l)$ of $\{A,B,C,D\}$ (hence with $i \neq j \neq k \neq l$). These must furthermore be such that the ``reduced'' matrices $W_{(i,j,k,l)} \coloneqq \Tr_c W_{(i,j,k,l),c}$ (where $c$ refers to the space of the final measurement), $W_{(i,j,k)} \coloneqq \Tr_l W_{(i,j,k,l)}$ (where $\Tr_l$ corresponds to the partial trace over the input and output spaces of the operation $U_l$), $W_{(i,j)} \coloneqq \sum_k \Tr_k W_{(i,j,k)}$ and $W_{(i)} \coloneqq \sum_j \Tr_j W_{(i,j)}$ are of the form
\begin{align}
W_{(i,j,k,l)} = \widetilde W_{(i,j,k,l)} \otimes \id^{l_O}, & \quad W_{(i,j,k)} = \widetilde W_{(i,j,k)} \otimes \id^{k_O}, \notag \\
W_{(i,j)} = \widetilde W_{(i,j)} \otimes \id^{j_O}, & \quad W_{(i)} = \widetilde W_{(i)} \otimes \id^{i_O} \label{eq:CCGO_constr}
\end{align}
for some matrices $\widetilde W_{(\cdots)}$ in the appropriate spaces. Here $\id^{l_O}$ denotes the identity operator on the output space of the operation $U_l$, and similarly for $\id^{k_O}$, $\id^{j_O}$ and $\id^{i_O}$.

To obtain the causal bound $p_\text{succ}^\text{CCGO}$ for all CCGO processes -- for a fixed choice of sets $\mathsf U_k$ and weights $q_k$, hence a fixed operator $G$ as defined in Eq.~\eqref{eq:Psucc_Tr_GW} -- one can then optimize the value of $p_\text{succ} = \Tr [G W]$ for all $W$ in the class described by Eqs.~\eqref{eq:CCGO_decomp}--\eqref{eq:CCGO_constr} above (which describes a closed convex cone, which we denote ${\cal W}^\text{CCGO}$) and with the additional normalisation condition \cite{Oreshkov2012,Oreshkov2016,Wechs2019} that $\Tr W = 2^4$:
\begin{align}
p_\text{succ}^\text{CCGO} = \ 
& \max_W \Tr \big[ G \, W \big] \notag \\
& \text{s.t.} \quad W \in {\cal W}^\text{CCGO},\ \Tr W = 2^4. \label{eq:SDP_primal}
\end{align}
As it turns out, this optimisation is a semidefinite programming (SDP) problem, which can in principle be solved faithfully \cite{Araujo2015,Branciard2016a,Wechs2019}.

Another possible, ``dual'' approach -- now just for a fixed choice of possible sets $\mathsf U_k$ -- is to optimize the causal witness rather than the process matrix. Fixing the witness to be of the form of $G$ in Eq.~\eqref{eq:Psucc_Tr_GW}, this allows us to optimize the weights $q_k$ for each $\mathsf U_k$: indeed the optimisation problem can be written here (see Appendix~H in Ref.~\citenum{Araujo2015}) as
\begin{align}
p_\text{succ}^\text{CCGO} = \ 
& \min_{p,\{q_k\}_k} p \notag \\
& \text{s.t.} \quad p\, \id/2^4 - G \in {\cal S}^\text{CCGO}, \notag \\
& \qquad G = {\textstyle \sum_k} \, q_k \, \KetBra{\mathsf U_k}^T \!\otimes \ketbra{y_k}{y_k}_c, \notag \\
& \qquad q_k \ge 0,\ {\textstyle \sum_k} q_k = 1, \label{eq:SDP_dual}
\end{align}
where 
\begin{equation}
	{\cal S}^\text{CCGO} \coloneqq ({\cal W}^\text{CCGO})^* \coloneqq \{S \mid \forall \, W{\in}{\cal W}^\text{CCGO},\ \Tr[SW] \ge 0 \}
\end{equation}
is the convex cone dual to ${\cal W}^\text{CCGO}$, which can, like the latter, be described in terms of SDP constraints \cite{Araujo2015,Branciard2016a,Wechs2019}.

Let us note here that the above characterisation of ${\cal W}^\text{CCGO}$ (via the decomposition of Eq.~\eqref{eq:CCGO_decomp}, with the matrices $W_{(\cdots)}$ satisfying the constraints of Eq.~\eqref{eq:CCGO_constr}) was shown \cite{Wechs2019} to be a sufficient condition for the process matrix to be ``causally separable'' \cite{Oreshkov2012,Oreshkov2016,Wechs2019}. It remains an open question, whether the class of causally separable processes is strictly larger than that of CCGO, or not. We nevertheless conjecture that the ``causal bounds'' $p_\text{succ}^\text{CCGO}$ we obtain here hold for all causally separable processes.

\subsubsection{Causal witnesses with finitely many settings}

As is clear from the discussion in App.~\ref{subsec:causal_attacks}, if one only uses the sets from Tables~\ref{tab:4setsU}a and \ref{tab:4setsU}b, then one can only get a trivial causal bound $p_\text{succ}^\text{CCGO} = 1$. In order to get a nontrivial bound, one needs to consider some other possible sets of unitaries.

To this end, we considered unitaries taken from the set
\begin{equation}\label{eqn:unitariesG}
  \mathcal G = {\textstyle \{\id, Z, X, Y, \frac{Z+X}{\sqrt2}, \frac{Z+Y}{\sqrt2}, \frac{X+Y}{\sqrt2}, \frac{\id+iZ}{\sqrt2}, \frac{\id+iX}{\sqrt2}, \frac{\id+iY}{\sqrt2}\} }
\end{equation}
(which have the nice property that their Choi matrices $\KetBra{U}$
span the full 10-dimensional space of all possible Choi matrices for qubit unitaries),
and looked for which sets $\mathsf U = \{U_A,U_B,U_C,U_D\}$ with $U_i\in \mathcal G$ satisfy the promise of Eq.~\eqref{eq:property_Hadamard}. We found 460 different such sets: 316 satisfying the promise for $y=0$, 60 for $y=1$, 42 for $y=2$ and again 42 for $y=3$.

The SDP problem of Eq.~\eqref{eq:SDP_dual} is large -- indeed, $G$ is a $2^{10} \times 2^{10}$ matrix and the characterisation of $\mathcal{S}^\text{CCGO}$ imposes many constraints -- making it at the limits of tractability.
To simplify the problem further, we exploit an approach based on ``quantum superinstruments'' introduced in Ref.~\cite{Wechs2020}.
To this end, we first note that Eq.~\eqref{eq:SDP_primal} can be simplified by rewriting Eq.~\eqref{eq:Psucc_Tr_GW} in the form
\begin{align}
& p_\text{succ} = \sum_y \Tr \big[ G^{[y]} \, W^{[y]} \big] \notag \\
& \quad \text{with} \quad G^{[y]} = {\textstyle\sum_k} \, \delta_{y,y_k} \, q_k \, \KetBra{\mathsf U_k}_c^T \notag \\
& \quad \text{and} \quad W^{[y]} = \Tr_c [(\id \otimes \ketbra{y}{y}_c) W], \label{eq:Psucc_Tr_GW_superinstr}
\end{align}
(where $\delta_{y,y_k}$ is the Kronecker delta).
Here, one now only needs to optimize over the four smaller $2^8 \times 2^8$ matrices $W^{[y]}$. 
The fact that the $W^{[y]}$'s must be obtained from some CCGO process as in the last line of Eq.~\eqref{eq:Psucc_Tr_GW_superinstr} above implies similar SDP constraints as Eqs.~\eqref{eq:CCGO_decomp}--\eqref{eq:CCGO_constr} on the $W^{[y]}$'s directly \cite{Wechs2020};
more formally, one has $\{W^{[y]}\}_y \in \overline{\mathcal{W}}^\text{CCGO}$ where $\overline{\mathcal{W}}^\text{CCGO}$ is again a closed convex cone.
The dual approach~\eqref{eq:SDP_dual} can then also be rewritten in the simpler form
\begin{align}
p_\text{succ}^\text{CCGO} = \ 
& \min_{p,\{q_k\}_k} p \notag \\
& \text{s.t.} \quad \{p\, \id/2^4 - G^{[y]}\}_y \in \overline{\mathcal{S}}^\text{CCGO}, \notag \\
& \qquad G^{[y]} = {\textstyle\sum_k} \, \delta_{y,y_k} \, q_k \, \KetBra{\mathsf U_k}_c^T, \notag \\
& \qquad q_k \ge 0,\ {\textstyle \sum_k} q_k = 1, \label{eq:SDP_dual_supertinstr}
\end{align}
where the dual cone
\begin{align}
	\overline{\cal S}^\text{CCGO} &\coloneqq (\overline{\cal W}^\text{CCGO})^* \notag\\
	 &\coloneqq \big\{\{S^{[y]}\}_y \ \big|\  \forall \, \{W^{[y]}\}_y{\in}\overline{\cal W}^\text{CCGO},\notag\\
	  & \qquad\qquad\qquad {\textstyle\sum_y}\Tr[S^{[y]}W^{[y]}] \ge 0 \big\}
\end{align}
can again be described by SDP constraints \cite{Wechs2020}.

We were able to solve the simpler SDP problem of Eq.~\eqref{eq:SDP_dual_supertinstr} using the 460 sets of unitaries from $\mathcal{G}$ with the solver SCS \cite{ODonoghue2016,scs_code}, obtaining a bound of $p_\text{succ}^\text{CCGO} \approx 0.92$.
We then progressively set to zero the smallest weights and solved the SDP again, eventually reaching 60 nonzero weights with no change in $p_\text{succ}^\text{CCGO}$ within numerical precision (36 corresponding to sets satisfying the promise for $y=0$, 12 for $y=1$, and 6 each for $y=2$ and $y=3$).

\subsubsection{Causal witnesses with random rotations}

\begin{table*}[!htbp]
\begin{tabular}{|r|c|c|c|c|c|c|c|c|c|c|c|c|c|c|c|c|c|c|c|c|c|c|c|c|c|c|c|c|c|c|}\hline
 $y$ 		& 0 & 0 & 0 & 0 & 0 & 0 & 0 & 0 & 0 & 0 & 0 & 0 & 0 & 0 & 0 & 0 & 1 & 1 & 1 & 1 & 1 & 1 & 2 & 2 & 2 & 2 & 3 & 3 & 3 & 3 \\ \hline
 $U_A$ & $\mathbb1$ & $\mathbb1$ & $\mathbb1$ & $Z$ & $\mathbb1$ & $\mathbb1$ & $Z$ & $\mathbb1$ & $Z$ & $Z$ & $\mathbb1$ & $Z$ & $Z$ & $Z$ & $Z$ & $Z$ & $\mathbb1$ & $Z$ & $Z$ & $Z$ & $Z$ & $Z$ & $\mathbb1$ & $Z$ & $Z$ & $Z$ & $\mathbb1$ & $Z$ & $Z$ & $Z$ \\ \hline
 $U_B$ & $\mathbb1$ & $\mathbb1$ & $Z$ & $\mathbb1$ & $\mathbb1$ & $Z$ & $\mathbb1$ & $Z$ & $\mathbb1$ & $Z$ & $Z$ & $\mathbb1$ & $Z$ & $Z$ & $Z$ & $X$ & $Z$ & $\mathbb1$ & $Z$ & $X$ & $X$ & $X$ & $Z$ & $\mathbb1$ & $X$ & $Z$ & $\mathbb1$ & $X$ & $X$ & $X$ \\ \hline
 $U_C$ & $\mathbb1$ & $Z$ & $\mathbb1$ & $\mathbb1$ & $Z$ & $\mathbb1$ & $\mathbb1$ & $Z$ & $Z$ & $\mathbb1$ & $Z$ & $Z$ & $\mathbb1$ & $Z$ & $Z$ & $X$ & $X$ & $\mathbb1$ & $X$ & $Z$ & $X$ & $Y$ & $X$ & $Z$ & $\mathbb1$ & $X$ & $Z$ & $\mathbb1$ & $Z$ & $Y$ \\ \hline
 $U_D$ & $Z$ & $\mathbb1$ & $\mathbb1$ & $\mathbb1$ & $Z$ & $Z$ & $Z$ & $\mathbb1$ & $\mathbb1$ & $\mathbb1$ & $Z$ & $Z$ & $Z$ & $\mathbb1$ & $Z$ & $Z$ & $\mathbb1$ & $X$ & $X$ & $X$ & $Y$ & $Z$ & $Z$ & $X$ & $\mathbb1$ & $Y$ & $X$ & $X$ & $\mathbb1$ & $Y$ \\ \hline
\end{tabular}
\caption{Table of 30 sets $\mathsf U = \{U_A,U_B,U_C,U_D\}$ involving the identity $\mathbb1$ and the orthogonal Pauli operators $X$, $Y$, $Z$ only, satisfying the promise of Eq.~\eqref{eq:property_Hadamard} (for some value of $y$, indicated in the first row), for the gate permutations $\Sigma=\{ABCD, BADC, CBDA, DACB\}$ and the Hadamard matrix of Eq.~\eqref{eq:Had_mat}.}
\label{tab:30setsU}
\end{table*}

The causal strategies described in App.~\ref{subsec:causal_attacks} exploit knowledge of the basis that the unknown unitaries are defined in.
A possibility to obtain better bounds on $p_\text{succ}^\text{CCGO}$ is therefore to allow the Verifier to provide the unitaries in an unknown basis.
Given a set $\mathsf{U}=\{U_A,U_B,U_C,U_D\}$, this corresponds formally to providing the operations $\mathsf{U}^{(V)}=\{V U_A V^\dagger, V U_B V^\dagger, V U_C V^\dagger, V U_D V^\dagger\}$ for some unknown unitary $V$. Note that if $\mathsf{U}$ obeys the promise of Eq.~\eqref{eq:property_Hadamard} then so does $\mathsf{U}^{(V)}$.

To construct better causal witnesses from this approach, we start as before with a fixed choice of sets $\mathsf{U}_k$ and then, in addition to choosing $\mathsf{U}_k$ with prior probability $q_k$, we randomly choose an unknown unitary $V$ to be applied according to the Haar measure.
Eq.~\eqref{eq:Psucc_Tr_GW} then becomes
\begin{align}
& p_\text{succ} = \Tr \big[ G \, W \big] \notag \\
& \quad \text{with} \quad G = \sum_k \, q_k \, \int \mathrm{d}\mu(V) \KetBra{\mathsf U_k^{(V)}}^T \!\otimes \ketbra{y_k}{y_k}_c, \label{eq:Psucc_Tr_GW_Haar}
\end{align}
where $\mu(V)$ is the normalized Haar measure over $\mathrm{SU}(2)$.
The SDP problems~\eqref{eq:SDP_primal}, \eqref{eq:SDP_dual} and \eqref{eq:Psucc_Tr_GW_superinstr} can then be solved in the same way as described above.

We again consider the 460 sets of unitaries $\mathsf{U}$ with each $U_i\in\mathcal{G}$ as in the previous section.
The integration over the Haar measure can be performed analytically by taking an explicit parameterisation of $\mathrm{SU}(2)$ unitaries.
However, since the $\KetBra{\mathsf U_k^{(V)}}^T$ are $2^8\times 2^8$ matrices, this procedure is nevertheless slow, even with automated symbolic integration using, e.g., Mathematica.
To simplify matters, we exploit that fact that the Haar measure is unitary invariant (i.e., $\mathrm{d}(V) = \mathrm{d}(UV) = \mathrm{d}(VU)$ for any unitary $U$), so sets $\mathsf{U}$ and $\mathsf{U}'$ that are equivalent up to a change of basis give $\int \mathrm{d}\mu(V) \KetBra{\mathsf U^{(V)}} = \int \mathrm{d}\mu(V) \KetBra{\mathsf U'_k{}^{(V)}}$.
We thereby find that there are 98 sets $\mathsf U$ which are in inequivalent in this way and which satisfy one of the properties $y_k$.

Considering witnesses constructed from these sets, we solved the dual SDP problem given in Eq.~\eqref{eq:SDP_dual_supertinstr}.
For CCGO processes, we find the bound $p_\text{succ}^\text{CCGO} \approx 0.84$.
Interestingly, we find that the same bound can be reached by considering the Haar randomisation only over witnesses using sets $\mathsf{U}$ containing only Pauli matrices, rather than from the full set $\mathcal{G}$.
Indeed, this bound can be obtained by randomising over the 30 sets $\mathsf U$ given in Table~\ref{tab:30setsU} that were found to have non-zero weights in the optimal witness we obtained.

\subsubsection{Derandomization}

In order not to require the assumption that the Prover does not know in which basis the Verifier provided each set $\mathsf U$, one could derandomize the approach above by using a weighted quantum $t$-design \cite{Gross2007}. This is a finite set of unitaries $X$ together with weights $\mathsf w$ such that the average of any operator over them is equal to its average over the Haar measure up to order $t$. Since $\Ket{\mathsf U_k^{(V)}}$ is an 8th order expression on $V$, an 8-design allows us to reproduce exactly the witness with bound $p_\text{succ}^\text{CCGO} \approx 0.84$ with a finite, fixed set of unitaries. Unfortunately, $t$-designs are rather large. It can be shown that the smallest size $|X|$ of a weighted 8-design for unitaries of dimension $2$ is bounded by $165 \le |X| \le 968$, so the resulting witness would have at least 4950 settings, and therefore be of little relevance for experiments \cite{Roy2009}.

In order to obtain smaller witnesses, we instead sampled 5 random qubit unitaries from the Haar measure, and conjugated all 30 columns of Table~\ref{tab:30setsU} with these fixed unitaries, obtaining a witness using 150 settings.

Solving the SDP in Eq.~\eqref{eq:SDP_primal} with the solver SCS, fixing the prior probability of choosing each set $\mathsf U_k^{(V)}$ to be $q_k/5$, where $q_k$ is the optimal weight obtained for the full randomization of $\mathsf{U}_k$ in the previous section, we obtained $p_\text{succ}^\text{CCGO}\approx 0.96$. 
By further optimising over all 150 weights using the dual SDP, we found this could be improved to $p_\text{succ}^\text{CCGO}\approx 0.93$.

By using more than 5 random unitaries this bound can be lowered further. For example, with 10 random unitaries we were able to obtain $p_\text{succ}^\text{CCGO}\approx 0.89$ (when optimizing over all $300$ weights).

\end{document}